\documentclass[11pt,a4paper]{article}
\usepackage{graphics}
\usepackage[english]{babel}
\usepackage{epsfig}

\newcommand{\JPsimm} {\mbox{\ensuremath{J/\psi \rightarrow \mu^+\mu^-}}}
\newcommand{\JPsiee} {\mbox{\ensuremath{J/\psi \rightarrow \mathrm{e}^+ \mathrm{e}^-}}}
\newcommand{\Psimm} {\mbox{\ensuremath{\psi(2S) \rightarrow \mu^+\mu^-}}} 
\newcommand{\Psiee} {\mbox{\ensuremath{\psi(2S) \rightarrow \mathrm{e}^+ \mathrm{e}^-}}}
\newcommand{\sqs} {\mbox{$\sqrt{\mathrm{s}}$}}

\newcommand{\ctsei           }{\mbox{CTEQ6m}}
\newcommand{\mrst            }{\mbox{MRST2002nlo}}
\newcommand{\ratio           }{\mbox{$\sigma_{\psi(2S)}/\sigma_{J/\psi}$}}
\newcommand{\rpsi            }{\mbox{$R_\psi$}}
\newcommand{\chisq           }{\mbox{$\chi^2$}}
\newcommand{\jpsi            }{\ensuremath{J/\psi}}
\newcommand{\primepsi        }{\ensuremath{\psi(2S)}}
\newcommand{\xfjpsi          }{\mbox{$x_F(J/\psi)$}}

\newcommand{\hb              }{\mbox{HERA-$B$}}

\newcommand{\sjpsi           }{\ensuremath{\sigma_{\jpsi}}}
\newcommand{\spsiprime       }{\ensuremath{\sigma_{\psi(2S)}}}
\newcommand{\lampsi           }{\ensuremath{\lambda_{\jpsi}}}
\newcommand{\lamprime       }{\ensuremath{\lambda_{\psi(2S)}}}
\newcommand{\etal           }{{\it et al.}}

\begin{document}

\begin{onecolumn}
\begin{flushright}
hep-ph/0601203 \\
25$^{\textrm{th}}$ January 2006\\
\end{flushright}

\vspace{0.5cm}
\begin{center}
\begin{LARGE}
{\bf Analysis of charmonium production at fixed-target experiments 
in the NRQCD approach}
\end{LARGE}
\end{center}
\vspace{0.5cm}
\begin{center}
F.~Maltoni,\\
\vspace{1mm}
\noindent
{\it
Institut de Physique Th\'eorique, Universit\'e Catholique de Louvain \\
Chemin du Cyclotron, 2, B-1348 Louvain-la-Neuve, Belgium} \\
\vspace{5mm}
J.~Spengler,\\
\vspace{1mm}
\noindent
{\it Max-Planck-Institut f\"ur Kernphysik, D-69117 Heidelberg, Germany \footnote{
Supported by the Bundesministerium f\"ur Bildung und Forschung, FRG, under contract numbers 05-7MP25I.}}
\vspace{5mm}

M.~Bargiotti,
A.~Bertin,
M.~Bruschi,
S.~De~Castro,
L.~Fabbri,
P.~Faccioli,
B.~Giacobbe,
F.~Grimaldi,
I.~Massa,
M.~Piccinini,
N.~Semprini-Cesari,
R.~Spighi,
M.~Villa,
A.~Vitale and 
A.~Zoccoli

\vspace{1mm}
\noindent
{\it Dipartimento di Fisica dell'Universit\`{a} di Bologna \\
and INFN Sezione di Bologna, I-40126 Bologna, Italy} \\

\vspace{5mm}
\noindent


\end{center}


\begin{abstract}

We present an analysis of the existing data on charmonium
hadro-production based on non-relativistic QCD (NRQCD) 
calculations at the next-to-leading order (NLO). 
All the data on \jpsi\ and \primepsi\ production in
fixed-target experiments and on $pp$ collisions at low energy are included.
We find that the amount of color octet contribution needed
to describe the data is about 1/10 of that found at the Tevatron.

\end{abstract}


\section{Introduction}
\label{sec:intro}

The production of charmonium and bottomonium states in high-energy
collisions has always been the subject of 
considerable interest~\cite{Brambilla:2004wf}. 
From the experimental point of view, charmonium decays into lepton pairs
offer very clean signatures that are used not only
for triggering and calibration but also to perform important physics
studies. The decay $B\to \jpsi+X$, for instance, provides
an easy handle to QCD studies of $b$ production and to the 
precise determination of some of the CKM parameters, such as $\sin 2\beta$. 

The importance of quarkonium is widely recognized also by the 
theoretical community. Charmonium and bottomonium states offer 
a unique laboratory for testing our understanding of QCD, 
and in particular of the interplay between the perturbative
and non-perturbative regimes, which describe the physics of 
heavy-quark creation and that of bound state formation, respectively.
The so-called color-singlet model~\cite{Berger:1981ni,Baier:1981uk} 
has been superseded by a rigorous framework, 
based on the use of non-relativistic QCD (NRQCD)~\cite{Bodwin:1994jh}, 
an effective field theory that consistently includes relativistic 
corrections and provides a solid ground for accurate theoretical analyses.

However, despite the theoretical developments and successes, 
not all the predictions of the NRQCD factorization approach 
have been firmly established. The first example is 
the universality of the non-pertubative matrix elements, on which
the predictive power of this approach relies. 
Measurements at
the Tevatron in proton-proton collisions suggest larger values for the
color-octet matrix elements than those obtained at HERA in 
electron-proton collisions~\cite{Kramer:2001hh}. 
Even more problematic is the measurement
of \jpsi\ polarization at the Tevatron. NRQCD predicts
a sizable transverse polarization for \jpsi 's at high-$p_T$, 
in contrast with the latest data that now clearly indicate 
that \jpsi 's are not transversely polarized~\cite{cdf-note}. 
While there is no quantitative and accepted explanation for this behaviour, 
it is generally argued that because the quarkonium mass is 
still not very large with respect to the QCD scale, in particular 
for the charmonium system, non-factorizable corrections  may not 
be suppressed enough  and/or the expansions in NRQCD may not converge 
very well. 

In the light of the present uncertain status, detailed studies on the
range of applicability of the NRQCD approach, above all for charmonium
states, are certainly welcome. In this work, we perform the NRQCD
analysis of charmonium production data as obtained from experiments 
at fixed-target (with the exclusion of that induced by pion beams) and
from $pp$ collisions at low energy.
Our purpose is twofold.  First, we present an up-to-date
collection of the experimental data on charmonium production in
fixed-target experiments. Second, we study whether data are consistent
with the NRQCD approach and in particular we extract information on
the color-octet contributions, to be compared with that obtained from
other experiments. 
Our analysis represents an improvement with respect to the previous studies
presented in Refs.~\cite{Beneke:1996tk,Maltoni:2000km}, both in terms of
accuracy of the theoretical predictions and for the more complete and
comprehensive treatment  of the experimental data.
The outline of the paper is as follows.  In the next
section, we briefly review the framework of the NRQCD approach and
state the theoretical results and assumptions that enter into our
predictions. In Section~\ref{sec:experiments} we discuss the
experimental data. In Section~\ref{sec:fit}  
we describe the fit strategy and present the results.
We draw our conclusions in the last section.

\section{The NRQCD approach}
\label{sec:nrqcd}

In the NRQCD  approach, 
the cross section for the production of a quarkonium
state $H$ in a nucleon-nucleon interaction is expressed as a sum of terms,
each of which factors into a short-distance coefficient and a
long-distance matrix element: 
\begin{eqnarray}
&&\sigma( pp \to H + X) =  
\sum_{i,j} \int dx_1 dx_2 f_{i/p} f_{j/p}\,
 \sum_n \hat \sigma(ij \to Q\overline{Q}\, [n] + x)\,
\langle {\cal{O}}^{H}\,[n]\rangle, 
\label{eq_fac}
\end{eqnarray} 
where the indexes $i,j$ run over all the partonic species and  $n$ denotes the color, 
spin and angular momentum state of an intermediate $Q\overline{Q}$ pair.  
The short-distance cross section
$\hat{\sigma}$ can be calculated as a perturbative expansion in the strong
coupling $\alpha_s$. The NRQCD matrix elements $\langle {\cal O}^H[n]
\rangle$ (see Ref.~\cite{Bodwin:1994jh} for their definition) are related
to the non-perturbative transition probabilities from the
$Q\overline{Q}$ state $n$ into the quarkonium $H$. They scale 
according to a definite power of the intrinsic heavy-quark velocity $v$ 
($v^2 \sim 0.3$ for charmonium and $v^2 \sim
0.1$ for bottomonium)~\cite{Lepage:1992tx}. 
The general expression (\ref{eq_fac}) is thus a double expansion in powers of $\alpha_s$ and $v$. 
While a formal and general proof of Eq.~(\ref{eq_fac}) is 
still lacking, it has been recently shown~\cite{Nayak:2005rt} that it holds for
high-$p_T$ quarkonium production up to its two-loop description.
In this work, we simply
assume that soft effects do not spoil factorization and Eq.~(\ref{eq_fac}) 
holds true also for total cross sections.

\begin{table}
\begin{center}
\vskip0.2cm
\renewcommand{\arraystretch}{1.5}
$$
\begin{array}{cccc}
\hline\hline
 H & \langle {\cal{O}}_1^{H} \rangle  & \langle
 {\cal{O}}_8^{H}[{}^3S_1] \rangle  & 
\langle {\cal O}_8^{H}[{}^1S_0^{(8)}]\rangle = 
\langle {\cal O}_8[{}^3P_0^{(8)}]\rangle/m_c^2\\ \hline
 \jpsi   & 1.16~{\rm GeV^3} & 1.19 \cdot 10^{-2}~{\rm GeV}^3 &  
 1.0 \cdot 10^{-2}~{\rm GeV}^3 \\[-1mm] 
 \primepsi & 0.76~{\rm GeV^3} & 0.50 \cdot 10^{-2}~{\rm GeV}^3 & 
 0.42 \cdot 10^{-2}~{\rm GeV}^3  \\[-1mm]
 \chi_{c0} & 0.11~{\rm GeV^5} & 0.31 \cdot 10^{-2}~{\rm GeV}^3 & --
 \\ \hline \hline
\end{array}
$$
\caption{Reference NRQCD matrix elements for charmonium production.
The color-singlet matrix elements are taken from the potential
model calculation of \cite{Buchmuller:1981su,Eichten:1995ch}. The
color-octet matrix elements have been extracted from the CDF data~\cite{Abe:1997yz} in Ref.~\cite{Nason:1999ta}.}
\label{tab:vevs}
\renewcommand{\arraystretch}{1.0}
\end{center}
\end{table}

The color-singlet short distance coefficients for 
spin singlet  $S$-wave ($^1S_0^{[1]}$), $P$-waves and all the leading
color-octet coefficients are known at NLO for both photon-proton 
and proton-proton collisions~\cite{Maltoni:1997pt,Petrelli:1997ge}. 
The color-singlet coefficient for $^3S_1^{[1]}$ is known at NLO only for
photon-proton collisions~\cite{Kramer:1995nb}. In this respect our analysis
cannot be considered as fully at NLO and should be updated once the
NLO calculation for the color singlet term will be available. On the other
hand, this is not an important limitation to our results, as will be made
clear in the following. 

The non-perturbative matrix elements have to be extracted from the data.
For the color-singlet terms this is straightforward. It can be easily
shown that, up to relativistic corrections of order $v^4$, they can be related
to those appearing in the corresponding decay rates and therefore can be extracted from
measurements of decay widths. On the other hand, color-octet matrix elements
can only be
extracted  from  production processes, such as photoproduction 
or hadroproduction.
The factorization hypothesis implies that the values extracted from different experiments
should be universal. Table~\ref{tab:vevs} shows the results of a fit performed on the 
CDF charmonium data~\cite{Abe:1997yz}, providing the leading color octet terms 
for $\jpsi,~\primepsi$ and $\chi_{cJ}$ production at the Tevatron energy~\cite{Nason:1999ta}. 
In the case of $S$-waves, the fact that transverse momentum distributions 
coming from CP-even states ($^1S_0^{[8]}$ and $^3P_J^{[8]}$) and the 
$^3S_1^{[8]}$ 
have different shapes, has been exploited to obtain information on their
relative size.   
An equally detailed information cannot be extracted from fixed-target results,
normally limited to the total production rates.
%

The analysis performed here is based on a code implementing the
NLO calculations of Ref.~\cite{Petrelli:1997ge}. For the theoretical 
inputs, we make the following choices. We use $\mu_0 = 2 m_c$  with $m_c=1.5$ 
GeV as our central value for  the renormalization, factorization and NRQCD scales 
(all taken equal). We estimate the associated uncertainty by varying 
the scales between $\mu_0$ and $4\mu_0$. The strong coupling constant
 $\alpha_S(m_Z)$ is tuned to the one used in the PDF sets, {\it i.e.},
\ctsei ~\cite{CTEQ6M} and \mrst ~\cite{MRST2002}.
We exploit spin symmetry to reduce the number of independent non-perturbative matrix elements,
\begin{eqnarray}
\langle {\cal O}_8^{\psi}(^3P_J) \rangle 
&=& (2 J + 1 )\;\langle  {\cal O}_8^{\psi} ({}^3P_0)  \rangle \ , \nonumber\\
\langle  {\cal O}_8^{\chi_{cJ} } (^3S_1) \rangle 
&=& (2 J + 1 )\;\langle  {\cal O}_8^{\chi_{c0} } ({}^3S_1) \rangle \ , \\
\langle  {\cal O}_1^{ \chi_{cJ} }(^3P_J) \rangle 
&=& (2 J + 1 )\; \langle {\cal O}_1^{ \chi_{c0} } ({}^3P_0) \rangle \nonumber
\end{eqnarray}
and consider only leading color octet corrections. 
We take the non-perturbative matrix elements  
collected in Tab.~\ref{tab:vevs} as our reference
values.~\footnote{
See also Ref.~\cite{Bodwin:2005hm} for a more recent analysis.}
Color-singlet matrix elements are kept fixed.
For the color-octet matrix elements we adopt the relative normalization
as that obtained from Tevatron measurements, but for the $S$-wave color-octet matrix-element  
two overall multiplicative numbers $\lambda_{\jpsi}$ and $\lambda_{\primepsi}$ are introduced
to be fitted with the fixed-target data.
The term $\langle {\cal O}_8^{\chi_{c0} } ({}^3S_1) \rangle$ is left fixed 
to its reference value extracted from the Tevatron data. 
With the above assumptions, the data are fitted with only two free parameters, 
the $\lambda$'s, which can be interpreted
as the fractions of the ``overall Tevatron octet contribution'' for $\jpsi$ and $\primepsi$
necessary  to explain the fixed-target data.

The following theoretical expressions for the cross sections are used:
\begin{eqnarray}
\label{eq:cross_jp}
&&\spsiprime =\sigma^{\rm D}_{\primepsi} \nonumber , \\
&&\sjpsi = \sigma^{\rm D}_{\jpsi}+
\sum_{J=0}^{2} {\rm Br}(\chi_{cJ} \to \jpsi \gamma) \,\sigma^{\rm D}_{\chi_{cJ}}+
{\rm Br}(\primepsi \to \jpsi X)\,\spsiprime , \\
&&\rpsi =\frac{\spsiprime}{\sjpsi} 
\nonumber\,,
\end{eqnarray}
where the superscript D refers to the direct contribution. 
For the branching ratios we use~\cite{PDG}
\begin{eqnarray}
{\rm Br}(\primepsi \to \jpsi X)&=& 57.6\%  \; \nonumber ,\\
{\rm Br}(\chi_{c0} \to \jpsi \gamma)&=& 1.18\% \; \nonumber ,\\
{\rm Br}(\chi_{c1} \to \jpsi \gamma)&=& 31.6\% \; , \\
{\rm Br}(\chi_{c2} \to \jpsi \gamma)&=& 20.2\% \nonumber .
\end{eqnarray}


\section{Present experimental situation} 
\label{sec:experiments}

The measurements of \jpsi\ hadroproduction have been performed in a time
period spanning about thirty years. Over such a long period, several different
experimental techniques have been used and different input information was
available at the time of the measurements. Therefore,
comparing results of different experiments on an equal footing
requires an update of the published numbers on several aspects.
For example, the charmonium branching ratios have changed with time
and the treatment of the nuclear effects are not homogeneous.
In our compilations, we update all the measurements with the
current best knowledge of branching ratios and nuclear effects.

\subsection{Compilation of {$\jpsi$}  cross sections}

The cross section for \jpsi\ production on a nuclear target of mass
number A is parametrized as  
\begin{equation}
\label{eq:pA}
   \sigma^{pA}_{J / \psi} =  \sjpsi  \cdot  A^{\alpha}.
\end{equation}
For this comparison the most precise
measurement of $\alpha = 0.96 \pm 0.01$~\cite{Leitch} at $\xfjpsi \simeq 0$ 
is used, assuming its independence of the cms-energy.
If an experiment has published cross sections obtained from different targets,
Eq.~(\ref{eq:pA}) is applied to obtain a combined result.
This is the case for the experiments CERN-PS~\cite{Bamberger} (H, C, W),
NA50~\cite{Alessandro} (Be, Al, Cu, Ag and W) and
\hb\ ~\cite{Abt} (C, Ti and W). As suggested in Ref.~\cite{Abreu2}, 
we combine the results of NA51~\cite{Abreu1} (H, D targets) and
NA38~\cite{Abreu2} (C, Al, Cu, W).

The results collected in Tab.~\ref{tab:sumexp} are updated
with the latest branching fractions~\cite{PDG} ($5.88 \pm 0.10 \%$
for \JPsimm\ and $5.93 \pm 0.10 \%$ for \JPsiee).
If only the forward cross section (\xfjpsi\ $>$ 0) is given, we
assume a symmetric $d\sjpsi/d x_F$  distribution and 
multiply the forward cross section by two to obtain the total cross section.
The cross sections of NA38, NA50 and NA51 
are only quoted for a limited phase-space.
In order to obtain total cross sections, the
extrapolation discussed in the NA51 paper is performed~\cite{Abreu1}.

Unless it is clearly stated in the publications, we assume that
the quoted uncertainties on branching ratio and atomic mass dependence were
taken into account in the systematic uncertainty. Therefore we consistently
update them in Tab.~\ref{tab:sumexp}.

The updated results for mid-rapidity and total cross sections are
displayed in Fig.~\ref{fig:sqsxsect}.

\begin{table*}[htb]
\begin{center}
\begin{tabular}{|l|c|c|c|c|}
\hline
Experiment                & Reaction & \sqs\  & 
$\frac{d\sjpsi}{dy}|_{y=0}$   &  $\sjpsi$  \\
         &  & (GeV)  & (nb/nucleon) & (nb/nucleon)                     
\\ \hline  \hline
CERN-PS~\cite{Bamberger} &  pA  & 6.8  &             & 0.65 $\pm$ 0.06
\\ \hline
WA39~\cite{Corden}       &  pp  & 8.7  &             & 2.4 $\pm$ 1.2
\\ \hline
IHEP~\cite{Antipov}      &  pBe & 11.5 & 16 $\pm $ 5.2 &  20 $\pm$ 5.2
\\ \hline
E331~\cite{Anderson}     &  pBe & 16.8 & 84 $\pm $ 20  & 122 $\pm$ 40
\\ \hline
NA3~\cite{Badier}        &  pPt & 16.8 &             &  80 $\pm$ 15
\\ \hline
NA3~\cite{Badier}        &  pPt & 19.4 &             & 110 $\pm$ 21
\\ \hline
NA3~\cite{Badier}        &  pp  & 19.4 &             & 124 $\pm$ 22
\\ \hline
E331~\cite{Branson}      &  pC  & 20.6 &             & 256 $\pm$ 30
\\ \hline
E444~\cite{Anderson2}    &  pC  & 20.6 &             & 166 $\pm$ 23
\\ \hline
ISR~\cite{Cobb}          &  pp  & 23.0 & 100 $\pm $ 77 &
\\ \hline
E705~\cite{Antoniazzi}   &  pLi & 23.8 &             & 267 $\pm$ 30
\\ \hline
UA6~\cite{Morel}         &  pp  & 24.3 & 104 $\pm $ 19 & 152 $\pm$ 20
\\ \hline
E288~\cite{Snyder}       &  pBe & 27.4 & 131 $\pm $ 33 & 204 $\pm$ 51
\\ \hline
E595~\cite{Siskind}      &  pFe & 27.4 & 187 $\pm $ 12 & 306 $\pm$ 18
\\ \hline
NA38/51~\cite{Abreu1,Abreu2}&  pA  & 29.1 & 169 $\pm $ 13 & 292 $\pm$ 64
\\ \hline
NA50~\cite{Alessandro}   &  pA  & 29.1 & 188 $\pm$ 14 & 325 $\pm$ 67
\\ \hline
ISR~\cite{Kourkoumelis}  &  pp  & 30   & 154 $\pm$ 42 &
\\ \hline
ISR~\cite{Clark}         &  pp  & 30.6 & 111 $\pm$ 30 &
\\ \hline
ISR~\cite{Cobb}          &  pp  & 31   & 142 $\pm$ 93 & 
\\ \hline
E672/706~\cite{Abramov}  &  pBe & 31.6 &             & 274  $\pm$ 60
\\ \hline
E771~\cite{Alexopoulos}  &  pSi & 38.8 & 202 $\pm$ 17 & 333  $\pm$ 25
\\ \hline
E789~\cite{Schub}        &  pAu & 38.8 & 170 $\pm$ 30 & 327  $\pm$ 56
\\ \hline
HERA-B~\cite{Abt}        &  pA  & 41.6 & 392 $\pm$ 51 & 663  $\pm$ 87
\\ \hline
ISR~\cite{Nagy}          &  pp  & 52   & 204 $\pm$ 85 & 716  $\pm$ 303
\\ \hline
ISR~\cite{Amaldi}        &  pp  & 52   & 216 $\pm$ 54 &
\\ \hline
ISR~\cite{Clark}         &  pp  & 52.4 & 185 $\pm$ 12 & 
\\ \hline
ISR~\cite{Kourkoumelis}  &  pp  & 53   & 229 $\pm$ 52 &
\\ \hline
ISR~\cite{Cobb}          &  pp  & 53   & 280 $\pm$ 161&
\\ \hline
ISR~\cite{Clark}         &  pp  & 62.7 & 172 $\pm$ 15 &   
\\ \hline
ISR~\cite{Cobb}          &  pp  & 63   & 538 $\pm$ 346&
\\ \hline
ISR~\cite{Kourkoumelis}  &  pp  & 63   & 250 $\pm$ 56 & 
\\ \hline
PHENIX~\cite{Adler}      &  pp  & 200  & 1051$\pm$ 255& 4000 $\pm$ 938
\\ \hline
\end{tabular}
\end{center}
\vspace*{0.2cm}
\caption{Updated differential ($d\sjpsi/dy$ at $y$=0)
and total ($\sjpsi$) production cross sections in
proton-induced interactions. The pA symbol in the second column 
indicates that the cross section value is obtained by fitting different
target materials.
\label{tab:sumexp}}
\end{table*}

\begin{figure}[htb]
\begin{center}
\resizebox{0.9\textwidth}{!}{%
\includegraphics{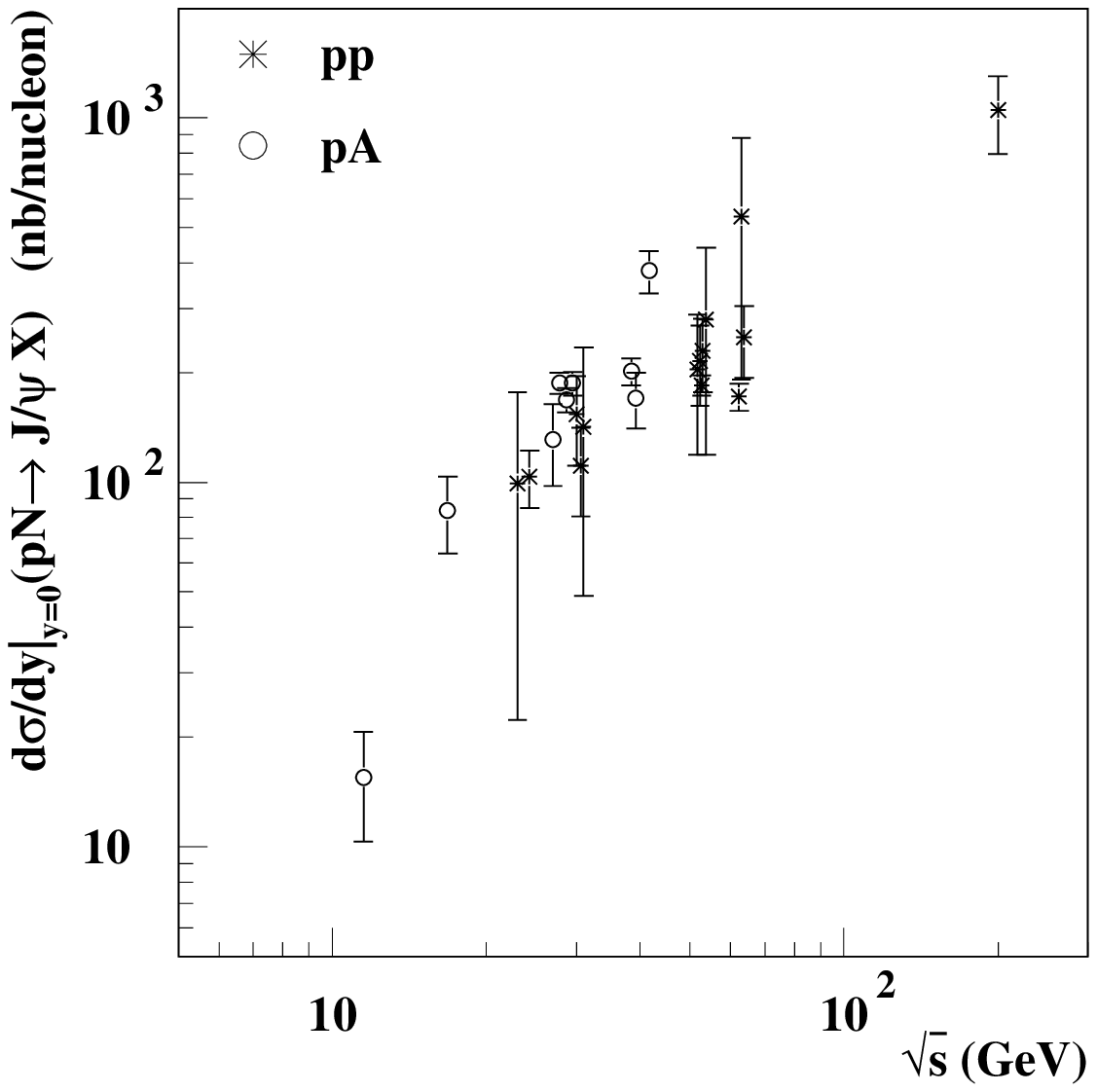}
\includegraphics{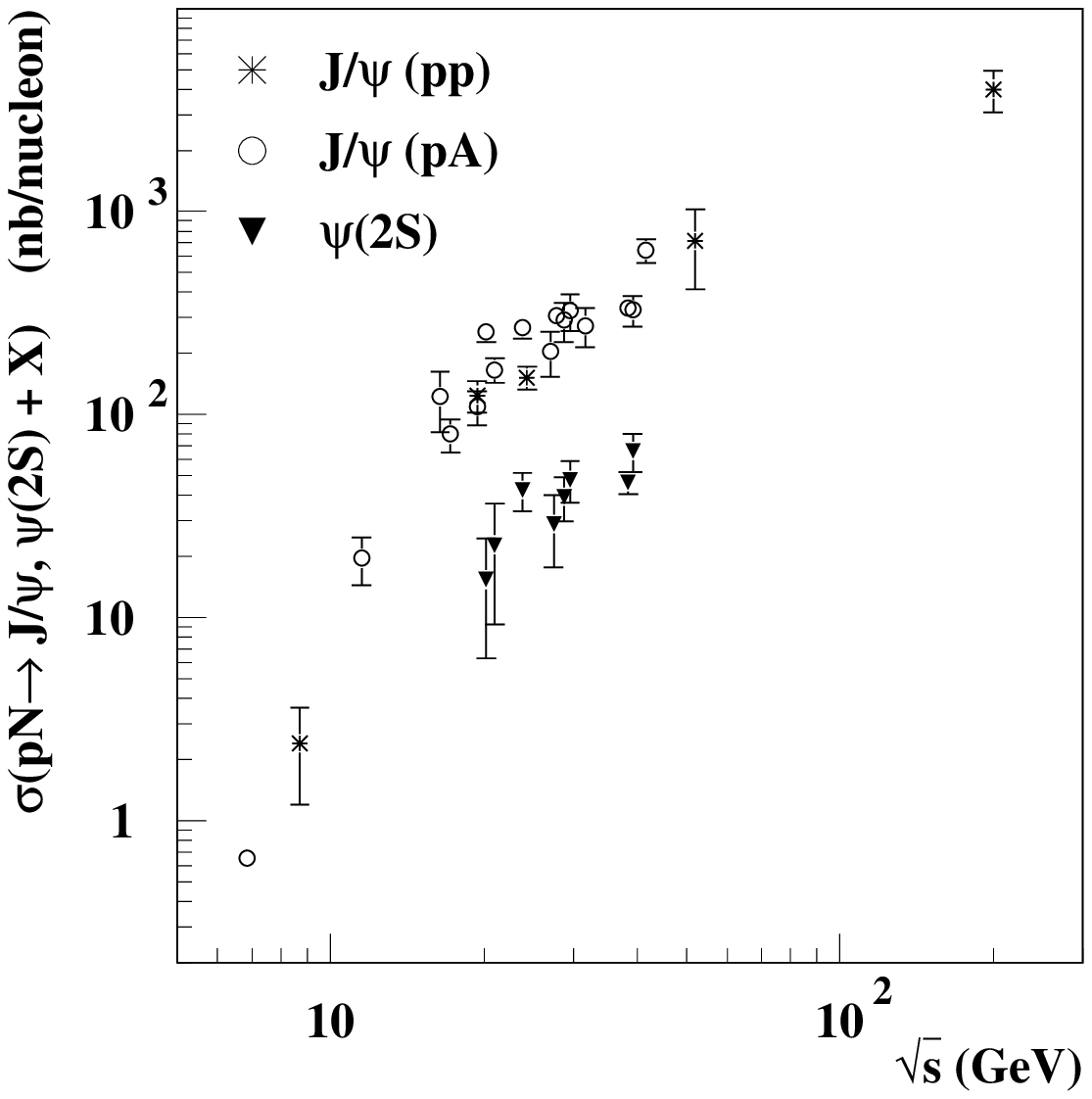}
}
\caption{Production cross sections in proton-induced interactions from
Tab.~\ref{tab:sumexp} as a function of the cms-energy.
Proton-proton (pp) and proton-nucleus (pA) measurements are indicated 
by different symbols.
Left: differential cross sections $d\sigma_{pN}/dy$ at $y$=0; 
right: total cross sections for \jpsi\  and \primepsi\  production.
\label{fig:sqsxsect}}
\end{center}
\end{figure}

\subsection{Compilation of $\primepsi$  cross sections}

The procedure described in the previous section is applied to the
published \primepsi\ cross sections using the
most recent branching fractions~\cite{PDG}  ($0.73\pm0.08 \%$
for \Psimm\ and $0.755\pm0.031 \%$ for \Psiee) and
$\alpha = 0.934\pm0.010$~\cite{Leitch} at $x_F(\psi(2S)) \simeq 0$.

The results on the 
\primepsi\ cross sections (see Fig.~\ref{fig:sqsxsect}) and
the ratios between \primepsi\ and \jpsi\ cross sections
are listed in  Tab.~\ref{tab:sumexp2}.

\begin{table*}[htb]
\begin{center}
\begin{tabular}{|l|c|c|c|c|}
\hline 
Experiment                & Reaction & \sqs\  & 
\spsiprime  &  \ratio  
\\
         &  & (GeV)  & (nb/nucleon) &  (\rpsi) 
\\ \hline  \hline
E331~\cite{Branson}      &  pC  & 20.6 & 15.4 $\pm$ 9.1  & 0.060 $\pm$ 0.035
\\ \hline
E444~\cite{Anderson2}    &  pC  & 20.6 & 22.8 $\pm$ 13.5 & 0.137 $\pm$ 0.079
\\ \hline
E705~\cite{Antoniazzi}   &  pLi & 23.8 & 42.5 $\pm$ 9.0  & 0.159 $\pm$ 0.029
\\ \hline
E288~\cite{Snyder}       &  pBe & 27.4 & 28.9 $\pm$ 11.3 & 0.141 $\pm$ 0.042
\\ \hline
NA38/51~\cite{Abreu1,Abreu2} &pA & 29.1& 39.3 $\pm$ 9.6 & 0.135 $\pm$ 0.015 
\\ \hline
NA50~\cite{Alessandro}   &  pA  & 29.1 & 47.1 $\pm$ 10.9 & 0.145 $\pm$ 0.017
\\ \hline
E771~\cite{Alexopoulos}  &  pSi & 38.8 & 46.3 $\pm$ 5.7  & 0.139 $\pm$ 0.020
\\ \hline
E789~\cite{Schub}        &  pAu & 38.8 & 66.1 $\pm$ 14.1 & 0.202 $\pm$ 0.055
\\ \hline
\end{tabular}
\end{center}
\vspace*{0.2cm}
\caption{Updated total production cross sections of \primepsi\
and ratios of \primepsi\ to \jpsi\ cross sections
in proton-induced interactions.  
\label{tab:sumexp2}}
\end{table*}

\subsection{Comments on the available data}

The \jpsi\ cross sections have been usually measured  on large samples by
triggering on dilepton decays. Therefore the most precise data have
uncertainties dominated by systematic errors.
The measurements show a good overall consistency,  even though some of the
results are hardly compatible. For instance, the two results at 20.6 GeV
(E331~\cite{Branson} and E444~\cite{Anderson2}) 
differ by  roughly $2\sigma$. The  E705~\cite{Antoniazzi} result
at 23.8 GeV exceeds the UA6~\cite{Morel} one at 24.3 GeV by more
than  $3\sigma$, in contrast with the expectation (and the general trend) 
that the cross section should increase with the cms-energy. This is an
indication that probably in some measurements the systematic uncertainties
coming from triggering effects or luminosity determination have been
underestimated.  

The \primepsi\ cross sections are usually estimated on the same data sample of
the \sjpsi , but are poorer in statistics (by a factor around 60). Therefore
they are less sensitive to the accurate determination of the systematic
effects. Within the quoted uncertainties, there is a good internal consistency
among the different measurements.

In the \rpsi\ cross section ratios, the systematic effects on luminosity or
trigger mainly cancel out and the final uncertainty is usually dominated by
the \primepsi\ statistics. The measurements are all compatible.

Since the measurements on the polarization are all compatible at $3\sigma$
with the assumption of no polarization (\cite{Abramov,Alexopoulos,Chang}),
possible effects of the polarization in the \jpsi\ or \primepsi\ decays are
always neglected and in the evaluation of the cross sections
the polarization is assumed to be negligible for both charmonium states. 
It should be noted that if any polarization is present
(expecially for \primepsi\ where a precise information is lacking) some
results might be affected by a systematic uncertainty larger than that quoted.

\section{Fit results}
\label{sec:fit}

The cross sections values obtained from NRQCD calculations at NLO,
Eqs.~(\ref{eq_fac}) and (\ref{eq:cross_jp}), have been used to fit the
experimental results summarized in
Tabs.~\ref{tab:sumexp},~\ref{tab:sumexp2}.

The ratio of the two color octet matrix elements for \jpsi\ and \primepsi\ 
production to the Tevatron ones (\lampsi\ and \lamprime\ respectively)
are used to fit the theoretical predictions to the experimental values.

Since the cross section depends on the product of the matrix
elements and the PDF's as shown in Eq.~(\ref{eq_fac}), 
a change in the PDF can influence the result of the fit on the two color octet
matrix elements. To estimate such effect we used 
two different PDF sets, the \ctsei ~\cite{CTEQ6M} 
and the \mrst ~\cite{MRST2002}.
\vskip 0.2cm

The data considered in the fit (Tabs.~\ref{tab:sumexp} and 
\ref{tab:sumexp2})
include 21 results on
the total \jpsi\ cross section (cms-energy range: 
[6.7 : 200] GeV), three \primepsi\ cross sections 
(NA50~\cite{Alessandro}, E771~\cite{Alexopoulos} and E789~\cite{Schub}; 
cms-energy range: [29.1 : 38.8] GeV) and five cross section
ratios \rpsi = \ratio\ (cms-energy range: [20.6 : 29.1] GeV), 
for a total of 29 experimental results. 
This set of data shows a good overall consistency, except for the 
few measurements of \sjpsi\ already mentioned in the previous section.

As a first step, the fit has been performed on all the 29 experimental
results using the \mrst\ and the \ctsei\ PDFs.
Since the charmonium production from singlet states strongly depends  
on the factorization ($\mu_F$) and renormalization ($\mu_R$) scales, 
we opted for determining the proper scale as an additional free parameter of
the fit.
The resulting optimal scale values are 
$\mu=1.5 \mu_0 $ for \mrst\ and $\mu =2.6 \mu_0$ for \ctsei ,
where $\mu = \mu_F = \mu_R$ and $\mu_0=2 m_c$ with $m_c$=1.5 GeV.
The fit results obtained with these scale values are reported in  
Tab.~\ref{tab:fitpdf}.

\begin{table*}[htb]
\begin{center}
\begin{tabular}{|l|c|c|}
  \hline
           & \mrst\ & \ctsei\  \\
\hline
$\mu_F=\mu_R$         & $1.5 \mu_0 $    & $2.6 \mu_0 $    \\
$\chi^2/d.o.f.$       & 114/27 & 170/27 \\
\lampsi               &  $0.089 \pm 0.013$   &  $0.211 \pm 0.027$   \\
\lamprime             &  $0.061 \pm 0.012$   &  $0.112 \pm 0.017$   \\
\hline
\end{tabular} 
\end{center}
\vspace*{0.2cm} \caption{Results of the fit performed using the
optimal scale factors for the \mrst\ and the \ctsei\ PDFs.
\label{tab:fitpdf}}
\end{table*} 

The \chisq\ of the fit is poor for both fits, since,
as previously discussed, the measurements of \sjpsi\ are not 
internally compatible. 
A reasonable explanation is that in some cases the experimental
uncertainty might have been underestimated. With this assumption, we increased
the uncertainties of the fit results by a scaling factor ($s$=2.05 and $s$=2.51 respectively),
obtained following the PDG prescriptions~\cite{PDG}.
The ratios of the matrix elements (\lampsi , \lamprime) 
provided by the fits differ by about a factor two
mainly because of the different scales used. 
Nevertheless, in both fits  the color octet matrix elements are found to be
much smaller ($1/5 - 1/10$) than those extracted at the 
Tevatron~\cite{Nason:1999ta}. 
Given that the \ctsei\ PDF shows a worse adaptation to the data, we
have decided to choose the \mrst\ PDF fit 
as our baseline fit, while the results from the \ctsei\ PDF 
have been considered in the systematic uncertainties determination.

The fit results as a function of the cms-energy are displayed in
Fig.~\ref{fig:resfit} for the \mrst\ PDF. 
In the top plot the \jpsi\ cross
section is shown. In the bottom-left plot the \primepsi\ cross section is
shown, while the \ratio\ ratio is displayed in the bottom-right plot. The
open circles in the two bottom plots represent the results
calculated from the published papers and not used in the fit (see
Tab.~\ref{tab:sumexp2}). 
The dot-dashed line indicates the NRQCD predictions without any octet
contribution in the charmonium production, while the solid line is
the result of the fit.
The two dashed lines are obtained by independently varying 
the scales $\mu_F$ and $\mu_R$ between  
$\mu_0$ and $4~\mu_0$. 

\begin{figure}[htb]
\begin{center}
\resizebox{0.9\textwidth}{!}{%
\includegraphics{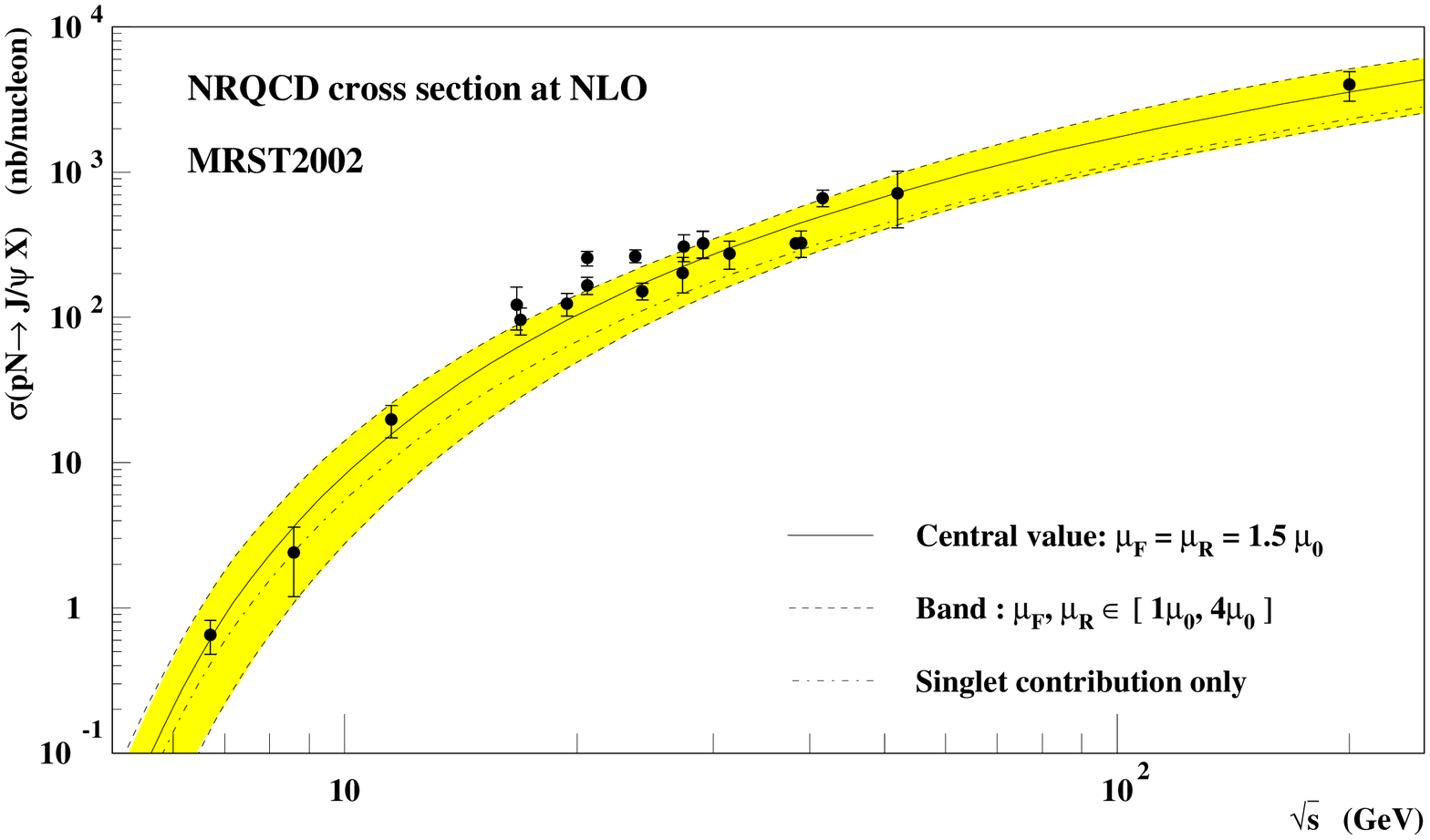} 
}
\resizebox{0.9\textwidth}{!}{%
\includegraphics{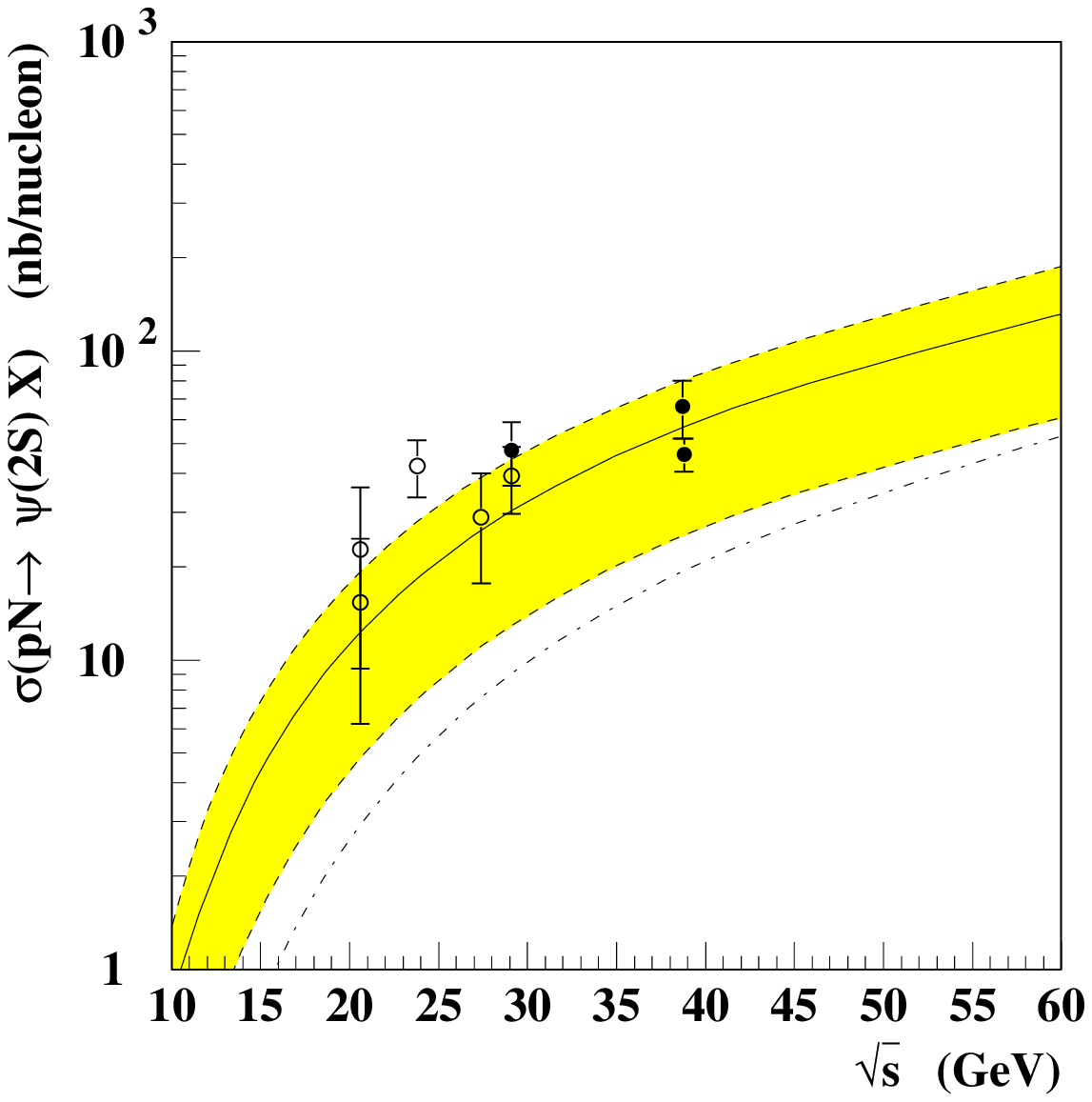}
\includegraphics{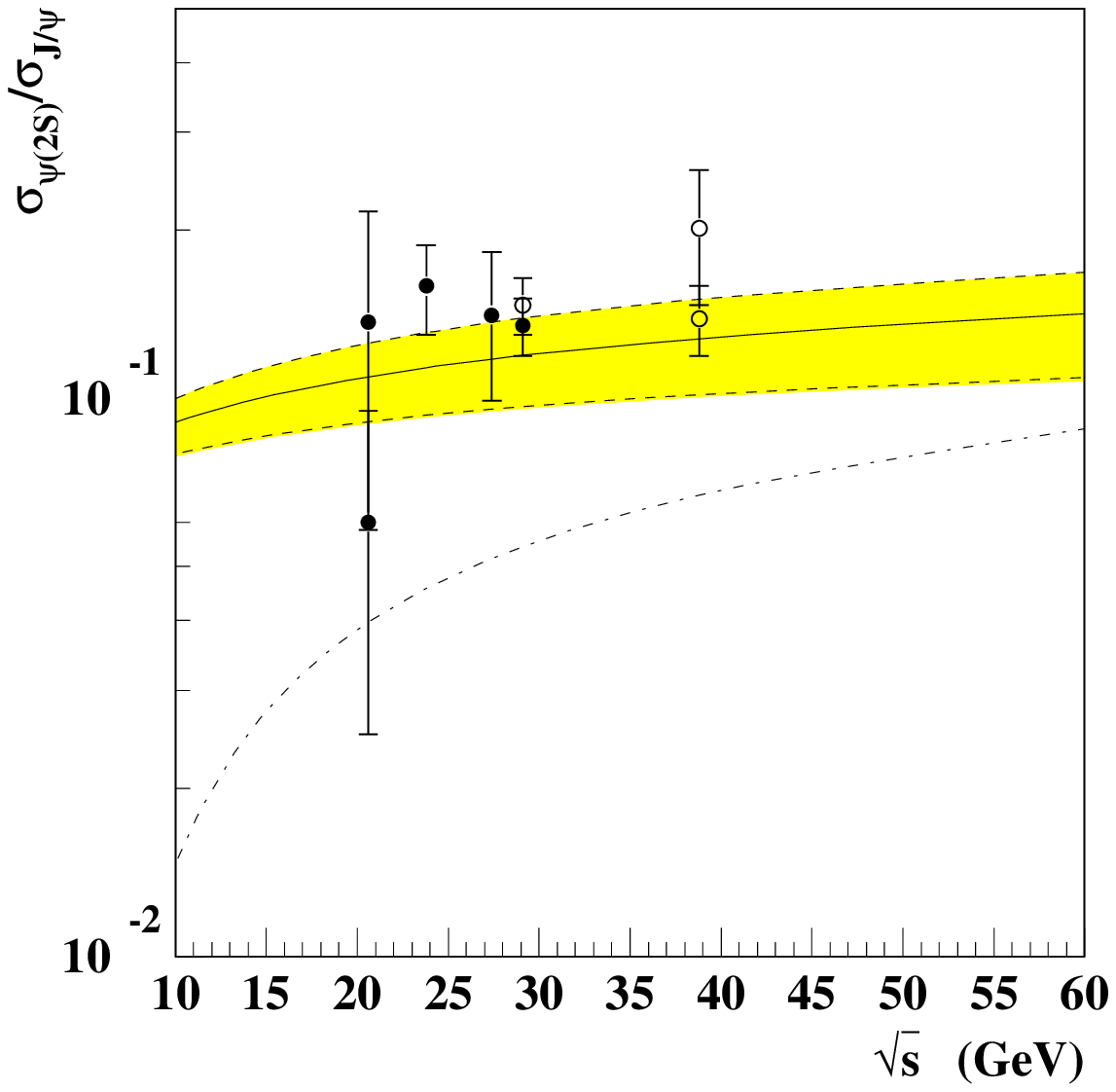}
}
\caption{Fit results as a function of the cms-energy for the
\jpsi\ cross section (top), the \primepsi\ cross section (bottom-left)
and the \ratio\ ratio (bottom-right). The open circles in the two
bottom plots represent the results calculated from the published
papers which are not used in the fit.
\label{fig:resfit}}
\end{center}
\end{figure}

\vskip 0.2cm

As one can see the curves obtained from the fit can well reproduce
the general trend of the \jpsi\ and \primepsi\ cross sections as a 
function of the cms-energy.
In order to fit the data the production from singlet states seems to
be not sufficient and contributions from octet states must be considered.
However, the fit shows that the amount of octet production  is small
compared to the results from Tevatron data~\cite{Nason:1999ta}
and is in agreement the results on charmonium production
obtained at HERA~\cite{Kramer:2001hh}, where
the octet contributions were negligible.


The stability of the results with respect to our baseline fit 
and the systematic uncertainties have been checked with detailed studies,
mainly by changing the selection of the measurements to be fitted and the PDF. 
First we excluded the experimental results which show a bad partial
\chisq\ ($\chi^2 _p>9$) and consequently are not compatible with the NRQCD
calculations.
Following these prescriptions the four \jpsi\ cross section measurements from the
E331~\cite{Branson}, E705~\cite{Antoniazzi}, E595~\cite{Siskind}
and E771~\cite{Alexopoulos} experiments were left out, 
obtaining a much improved \chisq\ ($\chi^2/dof = 37/23$),
with only a few percent variation on \lampsi\ and \lamprime .
Similar results have been obtained by excluding the four clearly incompatible 
measurements of \sjpsi\ at 20.6, 23.8 and 24.3 GeV (E331~\cite{Branson},
E444~\cite{Anderson2}, E705~\cite{Antoniazzi} and UA6~\cite{Morel})
discussed in the previous section. 
More fits have been performed by changing the cms-energy range of the
measurements and by using different target selections with the \mrst\ and the
\ctsei\ PDF. Again the results showed a very good stability on the cross
sections; the only significant change observed is, as shown in
Tab.~\ref{tab:fitpdf}, the dependence of the  \lampsi\ and \lamprime\
parameters on the actual scales ($\mu_F$ and $\mu_R$) and PDF used.
No significant variation of the results is observed when we include in the fit,
in addition to \sjpsi\ data, only \spsiprime\  or  only \ratio\ measurements.
In the quoted uncertainties, the correlations between different experimental
results, due to the dilepton decay branching ratios 
and to the luminosity and efficiencies determinations for different measures
performed by the same experiment, have been found to have a small impact.


\begin{figure}[tb]
\begin{center}
\resizebox{0.9\textwidth}{!}{%
\includegraphics{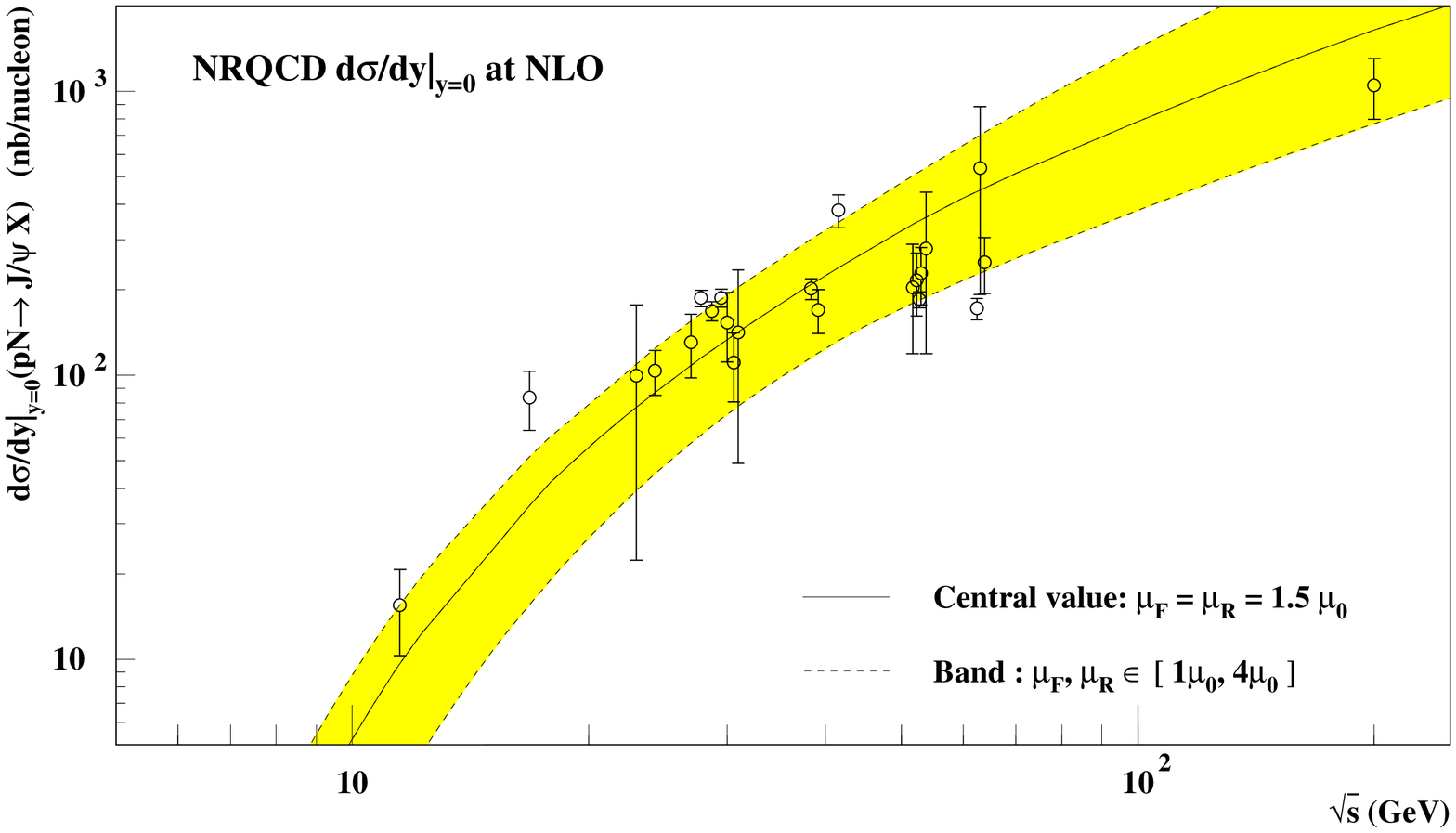} 
}
\caption{The \jpsi\ differential cross section $d\sigma_{pN}/dy$ at $y$=0,
as a function of the cms-energy. The data shown here are not used in the fit.
The theoretical prediction with its uncertainty 
corresponds to the fit to the total cross sections, as shown in
Fig.~\ref{fig:resfit}. 
\label{fig:resdsdy}}
\end{center}
\end{figure}

Once the stability of the fit procedure and its precision have been
established, one can obtain predictions for other quantities. 
As a first application, in Fig.~\ref{fig:resdsdy} we show the comparison
between the differential cross sections $d\sigma_{pN}/dy$ at $y$=0, as
a function of the cms-energy, and the theoretical predictions at NLO.
The data shown in the plot have not been used
in the fit and therefore this result is an important check of the
overall consistency of theory and experimental data.  As a further
application of our analysis, we present the predictions for the 
\jpsi\ and \primepsi\ cross sections and their ratio at $\sqrt{s}$ = 41.6 GeV, 
where \hb\ ~\cite{Abt} has
obtained the most recent result on the \jpsi\ cross section from a
Minimum Bias data sample.
The values obtained from the fit are the following:
\begin{equation}\label{eq:jp_hb}
    \sjpsi  = (502 \pm 44) \; \rm{nb/nucleon} \ ,
\end{equation}
\begin{equation}\label{eq:psi_hb}
    \spsiprime =  (65 \pm 11)\; \rm{nb/nucleon} \ ,\\
\end{equation}
\begin{equation}\label{eq:ratio_hb}
    \rpsi = (0.130 \pm 0.019) \ , \\
\end{equation}
where the quoted errors include the uncertainties due to the fit,
to the data selection and to the PDF. 
The systematic uncertainties due to the PDF have been obtained 
following the so called ``Les Houches Accord'' prescriptions 
as discussed in Ref.~\cite{LHA}.

\section{Conclusions}
\label{sec:conclusion}

In this paper we have collected all available data on \jpsi\ and
\primepsi\ production in hadron collisions (with the exception of data
obtained in pion-nucleus collisions) and updated them in the light of
the most precise determinations of nuclear effects.  We have then
presented their analysis in the context of NRQCD, using NLO predictions
for the short-distance cross sections and fitting the color-octet
non-perturbative matrix elements.  In order to ease the comparison
with the available determinations, we have chosen the values extracted
at the Tevatron as our reference. We find sizeable systematic uncertainties
associated both to the experimental data, which sometimes are
marginally consistent among themselves, and to the fixed-order nature
of theoretical predictions. Nevertheless, our results clearly indicate
that the amount of color-octet contributions needed to explain
fixed-target data is only about 10\% of that fitted at the Tevatron.
One can certainly argue that  part of this discrepancy 
might be associated to the fact that the Tevatron analysis is based 
on LO calculations only. On the other hand, the difference is
too large to be resolved by the inclusion of higher-order corrections.
In addition, it is plausible to speculate that once the NLO corrections to the 
color-singlet production were computed and included in
the analysis, there would be very little room left for color-octet
contributions to fit the fixed-target experiments, 
in close analogy to what happens in photoproduction~\cite{Kramer:2001hh} and
in agreement with the results in Ref.~\cite{khoze} where a different approach 
is used.


\section*{Acknowledgments}
\setcounter{footnote}{2}
\renewcommand\thefootnote{\fnsymbol{footnote}}

We would like to thank Mike Medinnis and many colleagues of the \hb\ Collaboration
for the support during this analysis and for many stimulating discussions.


\end{onecolumn}

\end{document}